# Investigations of pressurized Lu-N-H materials by using the hybrid functional

Wenfeng Wu,[1,2] Zhi Zeng,[1,2] and Xianlong Wang[1,2*]

[1] *Key Laboratory of Materials Physics, Institute of Solid State Physics, HFIPS, Chinese Academy of Sciences, Hefei 230031, China*
[2] *Science Island Branch of Graduate School, University of Science and Technology of China, Hefei 230026, China*

---

[*]Author to whom all correspondence should be addressed: xlwang@theory.issp.ac.cn

**Abstract**

Recently, Lu-N-H materials were reported to have room-temperature superconductivity, and the Hubbard $U$ correction on the Lu's $f$-electrons is necessary, and a constant $U$ = 5.5 eV was applied to different Lu-N-H configurations (Nature 615, 244 (2023)). Following simulations indicate that the superconducting transition temperature ($T_c$) of LuH$_3$ with $U$ = 0 eV is 50 ~ 60 K, while the N-doped LuH$_3$ is below 30 K. Quite recently, calculations with $U$ = 5 eV shows that the $T_c$ of N-doped LuH$_3$ exceeds 100 K. The properties of Lu-N-H are sensitive to the applied $U$ values. Here, the structural and electronic Lu-N-H properties at high-pressure (0 ~ 10 GPa) are systematically investigated based on the hybrid functional. We show that different Lu-N-H configurations should possess different $U$ values varying from 6.4 eV to 7.4 eV. Furthermore, at pressure ranging from 0 GPa to 1 GPa, the $f$ and $d$ band centers of N-doped LuH$_3$ show oscillation or even plateau, and the band gap of insulators also shows a platform near this pressure, this is consistent with the pressure range where room-temperature superconductivity appeared in Lu-N-H. Our work provides insights into the understanding of Lu-N-H materials and other hydrogen-rich superconductors based on the rare-earth elements.

**Keywords:** Lu-N-H, hybrid functional, correlation effects, electronic structure.

# 1. Introduction

Hydrides have gained significant attention in recent years due to their exceptional superconducting properties and promising applications. Of particular interest is the covalent hydride $H_3S$, which was predicted in 2014 [1] and experimentally demonstrated a superconducting transition temperature ($T_c$) of up to 203 K at 155 GPa. [2–3], making H-rich compounds the focus of many scientific investigations. A large number of H-rich compounds with ultra-high $T_c$ have been predicted and synthesized [4–14]. In 2019, the clathrate-like structure $LaH_{10}$, which is a rare-earth hydride, was reported to have a $T_c$ of 250 K at 130-220 GPa [9,15]. Very recently, an experimental study has discovered superconductivity in the Lu-N-H systems under near-ambient pressure (approximately 1 GPa), with an exceptional $T_c$ of 294 K [13]. This study pointed out that the room-temperature superconductor observed in the Lu-N-H systems can likely be attributed to the presence of cubic $LuH_{3-\delta}N_\varepsilon$ in the space group of $Fm\bar{3}m$, which is $LuH_3$ doped with N. This work has attracted widespread attentions [16–31].

The revelation of $LuH_{3-\delta}N_\varepsilon$ is a remarkable advancement in the field, and there is a great expectation of obtaining a thorough understanding of its physical mechanism, particularly in the areas of superconductivity and electronic structure properties under pressure. Sun *et al.* calculated electron-phonon coupling (EPC) strength $\lambda$ in the cubic $LuH_3$ phase and they found that there was no strong EPC was found in zone-center, regardless of whether the system was compressed or N-doped [31]. Ferreira *et al.* used theoretical calculations to determine that for the possible existence of Lu-N-H hydrides, with U = 0 eV, the $\lambda$ is one to two orders of magnitude lower than that of room-temperature superconductors and the $T_c$ N-doped $LuH_3$ is below 30 K [24]. In addition, by using temperature- and quantum-anharmonically-corrected phonon dispersions with the Migdal-Eliashberg formalism, Lucrezi *et al.* found that the $T_c$ for electron-phonon mediated superconductivity in cubic $LuH_3$ is in the range of 50 - 60 K [32], which is well below the reported room-temperature one [13]. These results suggest that the traditional electron-phonon-mediated pairing mechanism cannot be used to explain the superconductivity of the system. Guo *et al.* reported the finding of a magnetic transition at the temperature of about 56 K in N-doped lutetium hydrides, and the magnetic phase is robust against pressure up to 4.3 GPa [33]. Quite recently, Pavlov *et al.*'s work suggested that with $U$ = 5 eV, the T$_c$ of N-doped LuH$_3$ exceeds 100 K [34], The above results show that it is particularly important to consider electronic correlation effects.

Actually, in the work about 294 K superconductivity in the Lu-N-H systems [13], authors already pointed out that Lu's *f*-electrons play a critical role in determining the superconductivity and crystal structure of LuH$_{3-\delta}$N$_\varepsilon$, and ignoring their correlation effects could potentially result in inaccurate structural information [13]. Therefore, the properties of *f*-electrons should be necessary for understanding the superconductivity of Lu-N-H hydrides. The literature provides some possible Lu-N-H structures, i.e., LuH$_2$, LuH$_3$, LuH$_3$ with a N substituted for a H in an octahedral interstice (LuH$_3$-octN) and tetrahedral interstice (LuH$_3$-tetN), and a 2×2×2 supercell of the rhombohedral primitive cell of LuH$_3$ with a N substituted for a H in an octahedral interstice (pri-LuH$_3$-octN) and tetrahedral interstice (pri-LuH$_3$-tetN), and uses the GGA + $U$ ($U$ = 5.5 eV) method to describe the strong correlation effect of *f*-electron [13]. However, in the Hubbard $U$ correction method, $U$ is a semi-empirical value that is sensitive to the crystal structure [35,36]. Specifically, for Lu-N-H materials, due to different doping positions or concentrations of N, the chemical environments of Lu's *f* electrons are different, and the correlation effect of *f* electrons should be different. Obviously, using a constant $U$ value cannot describe Lu-N-H's electronic structural properties accurately.

In the recent theoretical simulations, the studies by Kim *et al.* and Pavlov *et al.* employed GGA + $U$ to account for the correlation effects [27], whereas most of the theoretical investigations directly studied Lu-N-H using the GGA method without considering the correlation effects present in Lu [24–26,31,32]. From this, it can be seen that the issue of applying the correct $U$ value on Lu-N-H materials has not been well solved, which will affect our understanding of the Lu-N-H system. The solution to this problem deepens our understanding of Lu-N-H and has important guiding significance for understanding the recently reported LuBeH$_8$. Li *et al.* predicted a new material, LuBeH$_8$, with superconducting temperatures up to 355 K at 100 GPa, and it indicated that the $f$-shell filled with lutetium hydride is expected to carry high $T_c$ [28].

Beyond the Hubbard $U$ correction, hybrid functional was successfully used in describing the strongly correlated systems [37–44], which do not need the semi-empirical determined $U$ values. For example, rare earth materials like lanthanide sesquioxides are known as strongly correlated materials that exhibit highly localized unpaired electrons in the $f$ band. Using the traditional GGA/LDA functional, inaccurate ground state and band gap predictions are obtained for $Ln_2O_3$ sesquioxides (Ln=La, Ce, Pr, Nd). Conversely, using a hybrid functional, the calculated structural parameters and band gap are more consistent with experimental data [44]. Our previous research has also demonstrated the ability of the hybrid functional to accurately describe the pressure-induced spin crossover behaviors of strongly correlated materials such as Fe-bearing MgO [45], $MnPS_3$ [46] and $CsV_3Sb_5$ [47]. It should be noted that, for the study of Lu-N-H materials, the hybrid functional method outperforms the GGA+$U$ method. This is due to two factors. Firstly, the hybrid functional method can be applied to all correlation systems, irrespective of external parameters, such as the value of Hubbard $U$. This means that the method is not dependent on the material under investigation. Secondly, near-room temperature superconductivity has been observed in the Lu-N-H system, often accompanied by variations in sample color, and the band gap is closely related to color. Notably, hybrid functional mixed with a portion of Hartree-Fock exact exchange is found to provide a more precise band gap than the GGA in most cases [48]. Therefore, in this study, we employ the HSE06 method in the hybrid functional to better understand the Lu-N-H materials.

In this work, hybrid functional is used to study $f$-band centers of Lu-N-H materials at 0 GPa. We find different Lu-N-H materials require different Hubbard $U$ to describe the $f$ electron correlation effect. Furthermore, the $f$ and $d$ band centers in most N-doped $LuH_3$ exhibit fluctuations or even a plateau near 1 GPa, and the band gap of insulators also shows a platform near this pressure, this is consistent with the pressure range where near-room-temperature superconductivity appeared in Lu-N-H. Our work reveals the novel behavior of correlated electrons and provides references for understanding near-ambient superconductivity.

## 2. Results and discussion

*2.1 Structural and electronic properties of Lu-N-H materials at 0 GPa*

Experimental evidence suggests that the Lu-N-H systems exhibit two distinct hydride compounds in almost all samples, which display superconductivity at 1 GPa. The chemical structure of both these compounds resembles that of face-centered cubic (FCC) metal sublattices, however, they contain varying amounts of hydrogen and nitrogen. These observations suggest that the system consists of N-doped cubic Lu hydride. The literature describes several possible compound structures, including $LuH_2$, $LuH_3$, $LuH_3$-octN, $LuH_3$-tetN, pri-$LuH_3$-octN, and pri-$LuH_3$-tetN [Figure S1]. For the system of octahedral and tetrahedral doping N in cubic $LuH_3$, the doping concentration is N : H = 1 : 11. Whereas in pri-$LuH_3$-octN and pri-$LuH_3$-tetN, which are 2 × 2 × 2 supercell of the rhombohedral primitive cell of $LuH_3$ with a N substituted for a H in an octahedral and tetrahedral interstice, N : H = 1 : 23.

The band center of *f* electrons in different Lu-N-H materials at 0 GPa will be discussed. From Table S1, we find that the *f*-band center of Lu in different Lu-N-H materials varies due to the difference in the chemical environment of Lu. In the cases of $LuH_2$ and $LuH_3$, the number of surrounding ligands H differs, leading to a difference of 1.582 eV in the *f*-band center. For the $LuH_3$ system, doping with N causes the *f*-band center to shift towards the Fermi level. Doping one N atom at the octahedral position in cubic $LuH_3$ results in a shift from -9.432 eV to -7.012 eV, a shift of 2.420 eV. Interestingly, the *f*-band center is lower when N is doped at the tetrahedral position than at the octahedral position, implying that octahedral doping has a more sensitive effect on the *f*-band center.

To evaluate the strength of the correlation interaction of Lu's *f* electrons, we determine the *U* values of various Lu-N-H systems using the *f*-band center obtained from the HSE06 calculations. We take $LuH_2$ at 0 GPa as an example. The *f*-band center of $LuH_2$ is -7.850 eV and we use different *U* values ranging from *U* = 5.5 eV to 7.0 eV with an interval of 0.2 eV to perform GGA + *U* calculations for the *f*-band center and obtained linear regression results [Figure 1(a)]. Finally, we find the $U$ value corresponding to -7.850 eV on the fitted line to be 6.354 eV (marked with a green pentagon). From this process, the $f$-band center shows a good linear relationship with the $U$ value. We use this method to determine the $U$ values of all the Lu-N-H systems at 0 GPa, and results are shown in the [Figure 1(b)]. We can find that different Lu-N-H materials possess different $U$ values, suggesting that the correlation strength of $f$ electrons cannot be sufficiently described by using a constant $U$ value. Among these materials, the smallest $U$ value is found in $LuH_2$, which is 6.354 eV, while the largest $U$ value is detected in pri-$LuH_3$-tetN, which is 7.400 eV. Additionally, it is noteworthy that the $U$ values of all Lu-N-H systems exceed the value of $U$ = 5.5 eV utilized in the existing literature [13].

Next, we study the N-doped $LuH_3$ system using the HSE06 method to reveal the bonding characteristics of N in $LuH_3$. The charge density distribution map reveals that N-doped $LuH_3$ is an ionic crystal [Figure S2]. In the case of octahedral N-doping, there is almost no electron cloud overlap between N and adjacent Lu atoms, while for tetrahedral doping, there is a slight overlap of electron clouds between N and Lu atoms, but without any characteristics of covalent bonding. The partial density of states (PDOS) is shown in Figure 2. For $LuH_2$ and $LuH_3$, the Fermi level is primarily contributed by Lu's $5d$ electrons. In $LuH_3$-octN, the Fermi level is mainly contributed by N's $p$ orbitals, while in pri-$LuH_3$-octN, it is mainly contributed by the $s$ orbitals of H atoms. This may be due to the reduced concentration of N doping, resulting in less contribution from N's $p$ orbitals near the Fermi level compared to $LuH_3$-octN. The PDOS of octahedral-doped $LuH_3$ crosses the Fermi level, indicating a metallic behavior, while the tetrahedral-doped $LuH_3$ remains an insulator. This implies that the different doping positions not only affect the $f$-band center but also have a significant impact on the electronic structure properties. Based on the HSE06 analysis, the band gaps for $LuH_3$-tetN and pri-$LuH_3$-tetN are 1.41 eV and 0.77 eV.

*2.2 Stability and electronic properties of Lu-N-H materials under pressure*

We first analyze the relative stabilities of LuH$_3$ with different N dopants under pressure by calculating the enthalpy. For LuH$_3$-octN and LuH$_3$-tetN [Figure S3 (a)], we take the enthalpy value of LuH$_3$-tetN under pressure as the reference point and found that the enthalpy differences between LuH$_3$-octN and LuH$_3$-tetN were positive within the considered pressure range, indicating that LuH$_3$-tetN is more stable than LuH$_3$-octN under 0 - 10 GPa. The enthalpy difference was 3.014 eV at 0 GPa and decreased with increasing pressure, reaching 2.466 eV at 10 GPa. For pri-LuH$_3$-octN and pri-LuH$_3$-tetN, the enthalpy value of pri-LuH$_3$-tetN was lower under all pressures [Figure S3 (b)]. These results suggest that the N-doped structure at the tetrahedral site is thermodynamically more stable and easier to synthesize experimentally for both doping concentrations considered.

The correlated electrons are important for the crystal and electronic structure of Lu-N-H materials. The *f*-band centers and *d*-band centers are shown in Figure 3 as a function of pressure. We first focus on how the *f*-band center evolves under pressure. For all the Lu-N-H systems discussed in this paper, the *f*-band center is generally on the rise. But around 1 GPa, the *f*-band centers of LuH$_3$ and LuH$_3$-octN show jumps at 0.2 GPa and 0.4 GPa, respectively. LuH$_3$-tetN's *f*-band center has experienced oscillating growth around 0 ~ 1 GPa. For pri-LuH$_3$-tetN, it remained constant between 0 GPa and 0.6 GPa and increased from 0.8 GPa. Previous studies have indicated that the strong correlation effects of *f* electrons have a significant impact on the structure of the Lu-N-H system. It is worth noting that in LuH$_3$-tetN, the center of its *f*-band undergoes significant changes at around 1 GPa, which may suggest that the structure of the system is sensitive near 1 GPa. However, for pri-LuH$_3$-octN, its *f*-band center increases linearly with pressure increasing.

Furthermore, we find that the center of the *d*-band shows a downward trend overall. This can be understood from the properties of the *d*-electrons. The *d*-electrons of Lu are not fully filled, and the *d* orbitals split into bonding and antibonding orbitals. After pressure loading, the splitting between bonding and antibonding orbitals becomes larger, and the center of the *d*-band below the Fermi level will shift downwards. Moreover, we find that the *d*-band centers in most Lu-N-H systems fluctuate irregularly near 1 GPa. However, we find that for LuH$_3$-octN and pri-LuH$_3$-tetN, *d*-band centers were insensitive to pressure at 0 ~ 1.0 GPa and 0 ~ 0.4 GPa, respectively.

For the two insulators LuH$_3$-tetN and pri-LuH$_3$-tetN, we investigate the band gap variation with pressure [Figure 4]. The results indicate that the band gap of LuH$_3$-tetN first increases to 1.42 eV at a pressure of 0.2 GPa, and then changes insignificantly until the pressure further increases to 2 GPa, when the bandgap begins to decrease. For pri-LuH$_3$-tetN, the band gap is in a plateau state in the range of 0 ~ 2 GPa. When the pressure is greater than 2 GPa, the band gap decreases with the increase in pressure.

## 3. Conclusion

In summary, we investigated the properties of correlated electrons in Lu-N-H materials using the hybrid functionals. First, at 0 GPa, the exact values of *U* in Lu-N-H materials are determined by using different *U* values to fit the *f*-band center, demonstrating that a single *U* value cannot describe the correlation effect in different Lu-N-H materials. Then, we studied the behavior of correlated electrons under pressure. The *f* and *d* band centers exhibited fluctuations or even plateaus around 1 GPa, and the band gap also showed a plateau at this pressure for LuH$_3$-tetN and pri-LuH$_3$-tetN. These phenomena occurred at the same pressure range as near-room-temperature superconductivity. Our theoretical work demonstrates that the hybrid functional is a promising approach to investigating Lu-N-H and other rare-earth-based hydrogen-rich superconductors.

## 4. Method

The first-principles calculation was performed utilizing the Vienna *ab initio* Simulation Package (VASP) [49,50]. All structures including the lattice constants and the atomic positions are fully relaxed based on the Heyd–Scuseria–Ernzerhof functional (HSE06) [48] with a separation parameter of 0.2 and a default mixing parameter of

AEXX = 0.25. The cutoff energy for plane wave expansion is set to 400 eV. The energy and force convergence thresholds are set to $1 \times 10^{-6}$ eV and 0.01 eV Å$^{-1}$. The lattice constants and internal atomic positions are fully relaxed. The Brillouin zone integral adopts the Monkhorst–Pack method and the k-point grid is set at $4 \times 4 \times 4$. Lu's $f$ orbitals are treated as valence orbitals. We calculate the *d/f* band center using the following formula,

$$d/f \text{ band center } = \frac{\int_{-\infty}^{\varepsilon_F} n(\varepsilon)\varepsilon \, d\varepsilon}{\int_{-\infty}^{\varepsilon_F} n(\varepsilon) \, d\varepsilon}$$

where $\varepsilon$ is the energy, and $n(\varepsilon)$ is the density of $d$ or $f$ electrons. $\varepsilon_F$ is the Fermi energy.

## Acknowledgments

The authors thank the support from the NSFC of China under Grant U2030114, and the CASHIPS Director's Fund No. YZJJ202207-CX. The calculations were partly performed in the Center for Computational Science of CASHIPS, the ScGrid of the Supercomputing Center and Computer Network Information Center of CAS, the CSRC computing facility, and Hefei Advanced Computing Center.

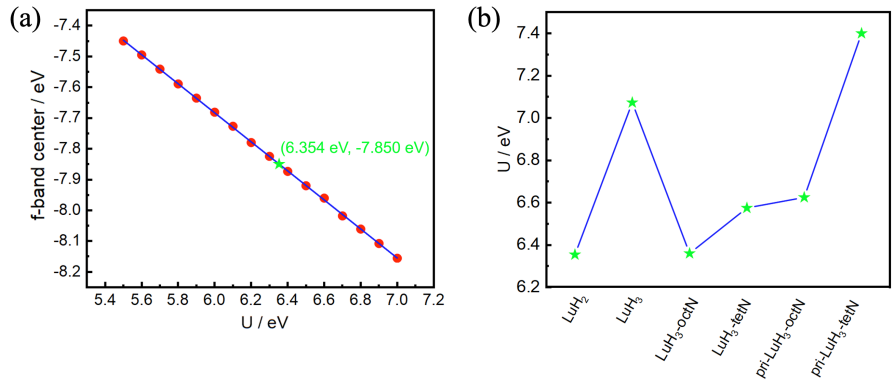

Figure. 1. (a) For LuH$_2$ at 0 GPa, the *f*-band center obtained from fitting a series of *U* values ranging from 5.5 eV to 7.0 eV is shown. The calculated *f*-band center of -7.850 eV using HSE06 is used to determine the *U* value of the system as 6.354 eV. (b) *U* values of various Lu-N-H materials at 0 GPa determined by combining the *U* fitting method with HSE06 calculation results.

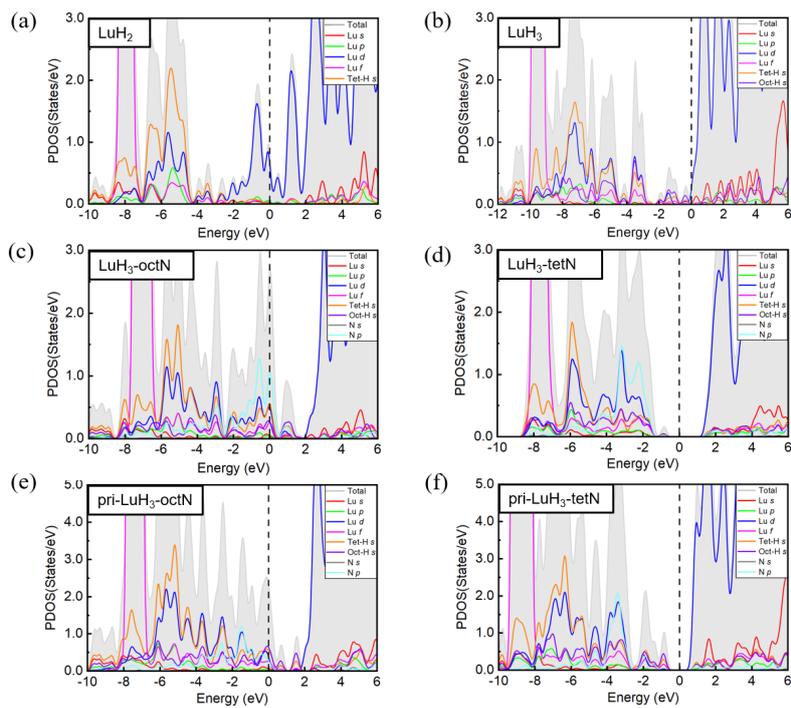

Figure 2. PDOS of Lu-N-H systems calculated using the HSE06 method at 0 GPa. (a)-(f) represent LuH$_2$, LuH$_3$, LuH$_3$-octN, LuH$_3$-tetN, pri-LuH$_3$-octN, and pri-LuH$_3$-tetN, respectively.

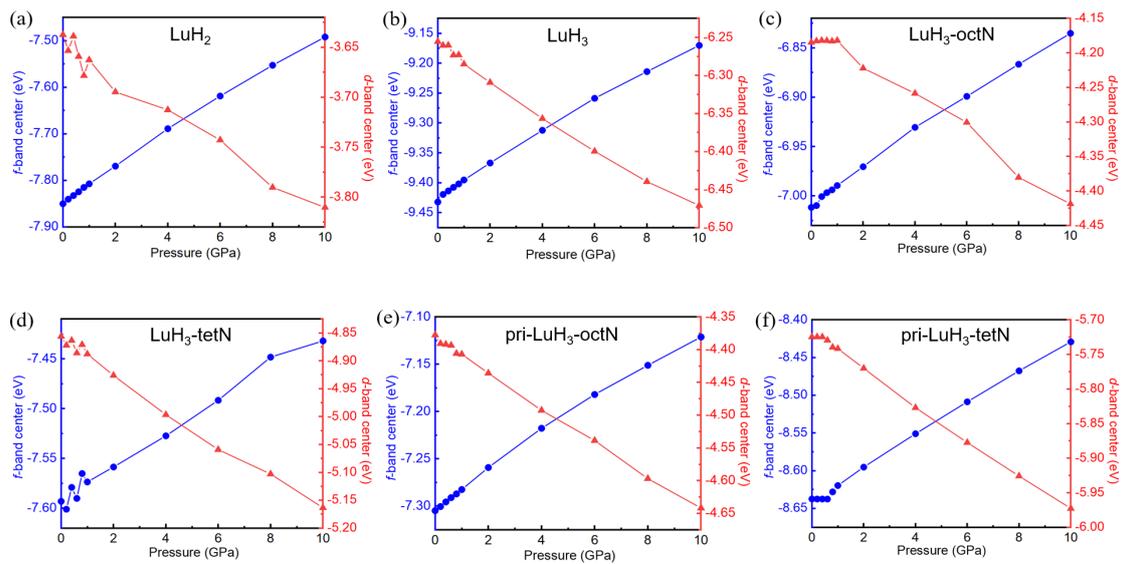

Figure. 3. The *f*-band centers and *d*-band centers of Lu atom in Lu-N-H systems under pressure by using the HSE06 method.

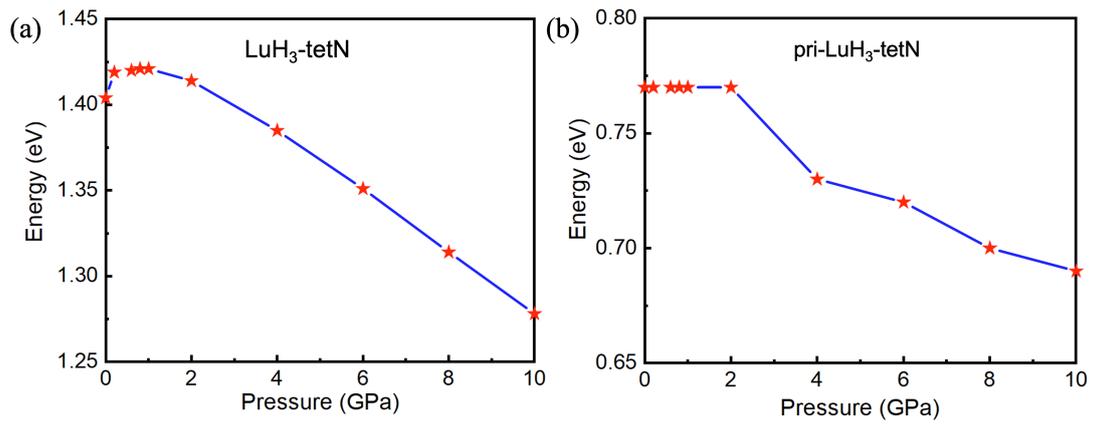

Figure. 4. The band gap of LuH$_3$-tetN and pri-LuH$_3$-tetN in different pressures by using the HSE06 method.